\documentclass[sigconf]{acmart}

\usepackage{xspace}		   
\usepackage{tikz}
\usetikzlibrary{trees}
\usetikzlibrary{shapes}
\usepackage{listings}
\usepackage{graphicx}
\usepackage{caption}
\usepackage{subcaption}
\usepackage{balance}

\usepackage{amsfonts}
\newcommand{\Y}{\checkmark}
\usepackage{pifont}
\newcommand{\N}{\hspace{1pt}\ding{55}}

\newcommand{\Tool}{{\sc AVclass2}\xspace}
\newcommand{\tool}{{\sc AVclass2}\xspace}
\newcommand{\avclass}{{\sc AVclass}\xspace}
\newcommand{\euphony}{{\sc Euphony}\xspace}

\newcommand{\superset}{Superset\xspace}

\newcommand{\token}[1]{{\em #1}}
\newcommand{\TAG}[1]{{\em #1}}
\newcommand{\IMPLIES}{{$\Rightarrow \,$}}

\newcommand{\numdatasets}{{11}\xspace}
\newcommand{\numsamples}{{42M samples}\xspace}

\renewcommand{\paragraph}{\vspace{3pt}\noindent\textbf}
\copyrightyear{2020}
\acmYear{2020}
\setcopyright{acmlicensed}
\acmConference[ACSAC 2020]{Annual Computer
Security Applications Conference}{December 7--11, 2020}{Austin, USA}
\acmBooktitle{Annual Computer Security Applications Conference (ACSAC
2020), December 7--11, 2020, Austin, USA}
\acmPrice{15.00}
\acmDOI{10.1145/3427228.3427261}
\acmISBN{978-1-4503-8858-0/20/12}

\acmBadgeR[https://www.acsac.org/2020/submissions/papers/artifacts/]{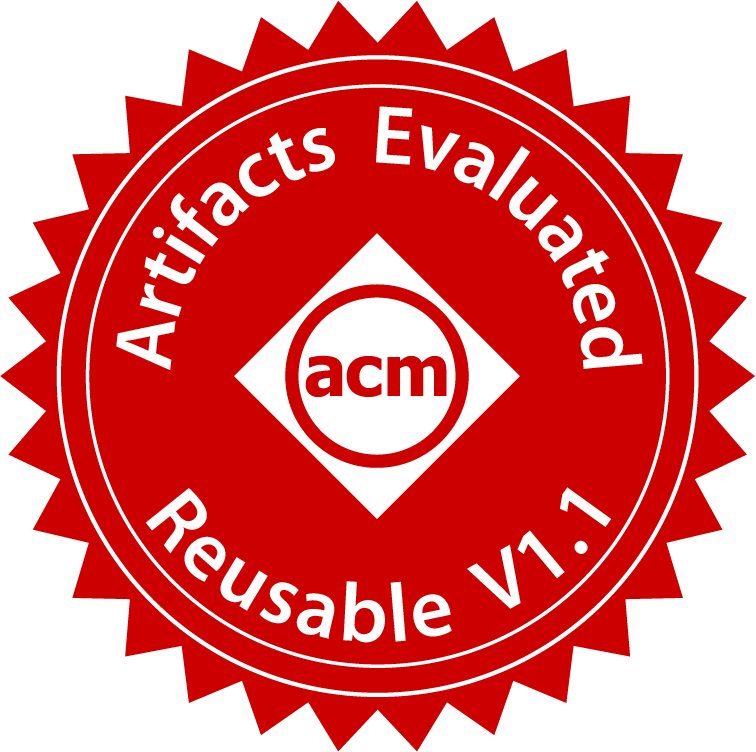}
\begin{document}

\title{\Tool: Massive Malware Tag Extraction from AV Labels}

\author{Silvia Sebasti{\'a}n}
\affiliation{  \institution{IMDEA Software Institute}
  \institution{Universidad Polit{\'e}cnica de Madrid}
}
\email{silvia.sebastian@imdea.org}

\author{Juan Caballero}
\affiliation{  \institution{IMDEA Software Institute}
  \city{Madrid, Spain}
}
\email{juan.caballero@imdea.org}

\begin{abstract}
Tags can be used by malware repositories and analysis services
to enable searches for samples of interest across different dimensions.
Automatically extracting tags from AV labels
is an efficient approach to categorize and index massive amounts of samples.
Recent tools like \avclass and \euphony
have demonstrated that, despite their noisy nature,
it is possible to extract family names from AV labels.
However, beyond the family name,
AV labels contain much valuable information
such as malware classes, file properties, and behaviors.

This work presents \tool, an automatic malware tagging tool
that given the AV labels for a potentially massive number of samples,
extracts clean tags that categorize the samples.
\Tool uses, and helps building, an \emph{open taxonomy} that organizes
concepts in AV labels, but is not constrained to a predefined set of tags.
To keep itself updated as AV vendors introduce new tags, it provides an 
update module that automatically identifies new taxonomy entries, 
as well as tagging and expansion rules that capture relations between tags. 
We have evaluated \tool on \numsamples and 
showed how it enables advanced malware searches and 
to maintain an updated knowledge base of malware concepts in AV labels.
  
\end{abstract}



\maketitle 

\section{Introduction}
\label{sec:intro}

Tags are keywords assigned to data objects
(e.g., documents, videos, images)
to categorize them and to enable efficient searches along different dimensions 
such as properties, ownership, and origin. 
Tags can be used by malware repositories and malware analysis services
(e.g.,~\cite{vt,joesandbox})
for enabling searches for samples of interest.
Malware tags can be manually produced by analysts, 
e.g., during reverse engineering of a malware sample, 
or output by analysis tools such as
sandboxes and signature-matching engines.
In both cases, each analyst and tool developer may use its own
\textit{vocabulary}, i.e., their own custom set of tags.
This is similar to user tagging, or \textit{folksonomies}, in Web 
services~\cite{halpin2007complex,korner2010stop},
which are known to lead to issues such as tags produced by different entities
being \textit{aliases} (or synonyms) for the same concept, 
some tags being highly specific to the entity producing them, and 
a tag from an entity corresponding to multiple tags from another entity.
To address these issues, standards such as 
Malware Attribute Enumeration and Characterization (MAEC)~\cite{maec} 
define a language for sharing malware analysis results.
However, they have low adoption due to their 
use of rigid \textit{controlled vocabularies} (i.e., predefined tags)
that may not always match analyst needs, 
require frequent updates, and
are necessarily incomplete, 
e.g., MAEC does not include malware families.

Detection labels by anti-virus engines (i.e., {\em AV labels}) 
are an instance of the above problem. 
An AV label can be seen as a serialization of the tags an AV engine 
assigns to the sample. 
Since tags are selected by each AV vendor rather independently, 
inconsistencies among labels from different vendors are widespread, 
as frequently observed in malware family names~\cite{bailey2007malware,canto2008large,maggi2011finding,avmeter,avclass,euphony}.
Recent tools like \avclass~\cite{avclass} and \euphony~\cite{euphony} 
demonstrate that, despite their noisy nature, 
it is possible to extract accurate family tags from AV labels. 
However, beyond the family name, 
AV labels may also contain much valuable information
such as the class of malware 
(e.g., \TAG{ransomware}, \TAG{downloader}, \TAG{adware}), 
file properties (e.g., \TAG{packed}, \TAG{themida}, \TAG{bundle}, \TAG{nsis})
and behaviors (e.g., \TAG{spam}, \TAG{ddos}, \TAG{infosteal}).

Automatically extracting malware tags from AV labels 
enables to cheaply categorize and index massive amounts of samples 
without waiting for those samples to be statically analyzed or 
executed in a sandbox.
And, since different AV vendors may execute a sample, 
the extracted tags may accumulate behaviors observed under 
different conditions. Furthermore, AV labels may encode domain knowledge from human analysts 
that is not produced by automated tools.
Once obtained, the tags can be used to  
enable efficient search of samples of a specific class, 
type, family, or with a specific behavior.
And, the identified samples can then be used as 
ground truth for machine learning 
approaches~\cite{bayer2009scalable,perdisci2010behavioral,malheur,drebin}.

Unfortunately, extracting useful tags from AV labels 
is challenging due to different vocabularies used by AV engines.
For example, Engine1 may include \TAG{multiplug} in its label for Sample1, 
while Engine2 includes \TAG{brappware} in its label for Sample2, 
which is not detected by Engine1. 
Both tags refer to samples that modify the browser by installing plugins. 
Thus, when an analyst searches for behavior \TAG{browsermodify}, 
both samples should appear as results.

This work presents \tool, an automatic malware tagging tool 
that given the AV labels for a potentially massive number of samples, 
extracts for each input sample a clean set of tags that capture properties 
such as the malware class, family, file properties, and behaviors.
\Tool ships with a default \emph{open taxonomy} that classifies 
concepts in AV labels into \emph{categories}, 
as well as default \emph{tagging rules} and \emph{expansion rules} 
that capture relations between tags.
In contrast to closed taxonomies, \tool does not mandate a predefined set 
of tags. 
Instead, unknown tags in AV labels, e.g., a new behavior or family name, 
are also considered.
\Tool has an \textit{update module} that 
uses tag co-occurrence to identify relations between tags. 
Those relations are a form of generalized knowledge that the
update module uses to automatically generate taxonomy updates, 
tagging, and expansion rules to keep the tool updated 
as AV vendors introduce new tags. 
Thus, \tool can maintain an updated knowledge base of malware concepts 
in AV labels.
 
\tool builds on \avclass~\cite{avclass}.
The goal is to perform the minimum needed changes to \avclass
that allow it to evolve from a malware labeling tool, 
focused exclusively on malware families, 
to a malware tag extraction tool that provides rich
threat intelligence by extracting and structuring all useful information 
in AV labels.
Thus, AVClass2 inherits AVClass major properties:
scalability,
AV engine independence,
platform-agnostic,
no access to samples required (only to their labels), and 
open source. 

We have evaluated \tool on \numsamples and compared it with 
\avclass and \euphony, the two state-of-the-art malware family labeling tools. 
We show how the tags \tool extract enable rich searches on malware samples, 
not possibly with existing tools, 
how the extracted tags are complementary to those already in 
use by popular malware repositories such as VirusTotal~\cite{vt}, and 
how the update module can be used to maintain
an updated knowledge base of malware concepts in AV labels.

The main properties of \tool are:

\begin{itemize}
  \item Automatically extracts tags from AV labels that 
    categorize malware samples according to their 
    malware class, family, behaviors, and file properties.

  \item Uses and builds an open taxonomy that does not use a 
    closed set of tags, 
    and thus can handle new tags introduced over time by AV vendors. 

  \item Can expand the input taxonomy, tagging rules, and expansion rules, 
    by generalizing relations found in AV labels, 
    allowing to maintain over time an up-to-date knowledge base of 
    malware concepts in AV labels.

  \item Evaluated on \numsamples and compared with the  
                two state-of-the-art malware family labeling tools~\cite{avclass,euphony}.

  \item Open source\footnote{https://github.com/malicialab/avclass.}.
   
\end{itemize}

\begin{figure}[t]
  \centering
  \includegraphics[width=\columnwidth]{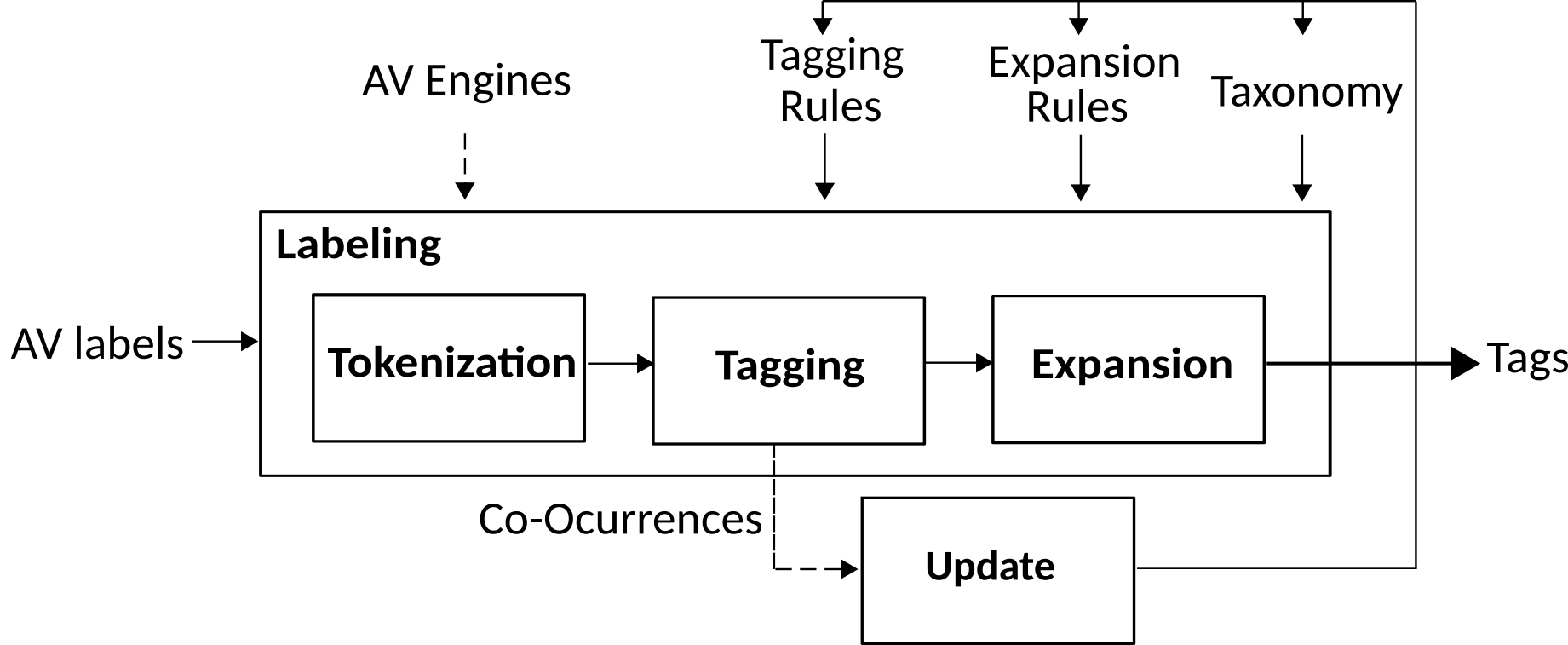}
  \caption{\tool architecture.} 
       \label{fig:arch}
\end{figure}

\section{Overview}
\label{sec:overview}

When an AV engine flags a sample as malicious 
it returns an {\em AV label} (or simply a {\em label} for short), 
i.e., a string with information about the malicious sample. 
An AV label can be seen as a sequence of tags separated by delimiters. 
Those tags are engine-specific as they are selected rather independently 
by the AV vendor.
For clarity, we will use \textit{tokens} to refer to engine-specific tags 
in AV labels and \textit{tags} to refer to entries in the taxonomy.
Some tokens in AV labels contain useful information such as the
malware class (e.g., \TAG{virus}, \TAG{worm}), 
malware family (e.g., \TAG{zbot}, \TAG{virut}), 
file properties (e.g., \TAG{pdf}, \TAG{toolbar}), 
and behaviors usually retrieved dynamically 
(e.g., \TAG{ddos}, \TAG{proxy}).
Other tokens do not provide useful information
because they are generic  (e.g., \token{malicious}, \token{application}) or 
very specific to the AV vendor, 
and thus of little use for external analysts, 
such as the detection technology (e.g., \token{deepscan}, \token{cloud}) or 
the specific variant of a malware family (e.g., \token{aghr}, \token{bcx}).
At a high level, the goal of \tool is to distinguish the tokens that provide 
useful information, identify relations between tokens from different AV 
engines, and convert them into tags in the taxonomy. 

The architecture of \tool is shown in Figure~\ref{fig:arch}. 
It comprises of two modules: 
{\em labeling} and {\em update}.
The labeling module takes as input the AV labels assigned by
multiple AV engines to the same samples, 
an optional list of AV engines whose labels to use
(if not provided, all AV labels are used),
a set of tagging rules, 
an optional set of expansion rules, and 
a taxonomy that classifies tags into categories and captures 
parent-child relationships between tags in the same category.
For each input sample, it outputs a set of tags
ranked by the number of AV engines 
including the tag's concept in their label.
\Tool ships with default tagging rules, 
default expansion rules, and a default taxonomy.
Thus, \tool can be used out-of-the-box without the need 
for any configuration. 
However, \tool is fully configurable, so the analyst can easily 
plug-in its own tagging rules, expansion rules, and taxonomy.

The update module takes as input the co-occurrence statistics,
tagging rules, expansion rules, and taxonomy. 
It first identifies strong relations between tags, 
which generalize knowledge beyond individual samples, 
e.g., that a family is ransomware or sends SMS. 
Then, it uses inference rules on the relations to 
automatically propose new tagging rules, new expansion rules, and 
taxonomy updates, which are then fed back to the labeling.

The remainder of this section introduces the labeling module 
in Section~\ref{sec:labelingOverview}, 
describes the input taxonomy in Section~\ref{sec:taxonomy}, and 
introduces the update module in Section~\ref{sec:update}.

\subsection{Labeling Module Overview}
\label{sec:labelingOverview}

The labeling module comprises of three steps: 
tokenization, tagging, and expansion. 
We detail them next.

\paragraph{Tokenization.}
The first step is to split each label into a list of tokens.
The key property of the tokenization is that it aims to be vendor-agnostic, 
i.e., it does not attempt to infer the structure of the labels used by each 
AV vendor. 
The reasons for this design decision are that 
the number of AV vendors is very large 
(e.g., over a hundred AV engines have been used by VirusTotal over time),
that each vendor provides different information in their AV labels, 
and that AV vendors are not very consistent in the format of their labels, 
often modifying them over time.
If we were to try to define, or automatically infer, templates for each 
label format, we would end up with hundreds of templates. 
This would make selecting the right template for a label challenging. 
Worse, whenever a label was observed using a previously unknown label format, 
tokenization errors could be introduced. 

\paragraph{Tagging.}
The main step in \tool is replacing a token in an AV label with a 
set of tags in the taxonomy.
Such tagging converts the unstructured 
information in the AV labels into well-defined concepts in the taxonomy. 
It also identifies potentially useful tokens not yet tagged. 
Tagging uses a set of input {\em tagging rules} 
that specify the set of tags that a token will be replaced with, 
i.e., a tagging rule maps an input token to an output set of tags. 
Most tagging rules map a token to a single tag. 
For example, a tagging rule maps the \token{downldr} token to the
\TAG{downloader} tag.
In this case, we say that that the token is an {\em alias} for the tag, 
since it captures the same concept as the tag. 
Multiple tokens may be aliases for the same tag.
For instance, another tagging rule maps the \token{dloader} token 
also to the \TAG{downloader} tag. 
Thus, tokens \token{downldr} and \token{dloader} are both 
\TAG{downloader} aliases.
Similarly, tokens \token{finloski} and \token{fynloski} are 
aliases for the \TAG{darkkomet} family tag.
Aliases enable identifying when different vendors use 
different tokens to capture the same concept. 
A tagging rule may also define that a token maps to 
multiple tags if the token captures multiple concepts.
For example, the token \token{ircbot} maps to tags 
\TAG{irc} and \TAG{bot}.
It is possible to define a tagging rule that maps a token to a tag 
that is identical to the token. 
For example, the \token{downloader} token maps to the \TAG{downloader} tag.
However, there is no need to define such rules as they are implicitly handled.
 
A tagging rule should map a token to the most specific tags possible. 
For example, a tagging rule could map token \token{addrop} to 
the \TAG{adware} tag, the \TAG{grayware} tag, or both of them. 
In this case, it is best to map it to the \TAG{adware} tag because 
the default taxonomy, detailed in Section~\ref{sec:taxonomy}, 
captures that \TAG{adware} is a child of \TAG{grayware}.
The expansion step below takes care of making explicit the implicit 
relationships between tags in the taxonomy. 
Thus, there is no need to assign 
the \TAG{grayware} tag explicitly at this point. 

A tagging rule can also define that a token maps to an empty set of tags. 
We call such tokens {\em generic} because they do not provide 
useful information. 
Generic tokens are discarded during tagging. 
Example generic tokens are \token{malicious} and \token{application}.
It is important to highlight that generic tokens in \tool are different from 
generic tokens in the original \avclass. 
In \avclass, a generic token provided no information about 
the family a sample belongs to. 
In \tool, a generic token provides no useful information at all 
(family or other). 
Thus, many tokens that were considered generic by \avclass will be assigned 
tags by \tool, e.g., \token{dldr} was a generic token in \avclass, 
but produces tag \TAG{downloader} in \tool.
In fact, generic tokens in \avclass are used as seeds
to build the default taxonomy for \tool, 
as explained in Section~\ref{sec:taxonomy}.

When a token has not been seen before, it will not have a tagging rule 
defined for it. 
In this case, the token should not be discarded because it 
may still provide useful information, 
e.g., it may capture a newly discovered family.
Thus, unknown tokens are passed to the expansion step.

\paragraph{Expansion.}
The expansion step uses {\em expansion rules} 
that define that a tag implies a set of other tags, 
where the original tag is included in the expanded set.
For example, the \TAG{worm} class tag can be expanded to 
include the \TAG{selfpropagate} behavior tag, 
since a worm self-propagates by definition.
Expansion rules allow to more accurately capture how many AV engines 
capture the same concept in their labels. 
For example, prior to expansion, \tool may output that the \TAG{worm} tag 
is assigned by two engines and \TAG{selfpropagate} by another three. 
After expansion, \tool can output that 
\TAG{selfpropagate} is assigned by five engines.
More importantly, expansion rules help increase search coverage. 
For example, if 95\% of samples with the \TAG{virut} family tag 
also have the \TAG{virus} class tag, 
the update module will output an expansion rule from \TAG{virut} to \TAG{virus}.
Using that expansion rule, an extra 5\% of \TAG{virut} samples 
will be added the \TAG{virus} tag and thus will show up in 
searches for virus samples.

There exist two types of expansion rules: 
{\em intra-category} and {\em inter-category}. 
Intra-category rules are implicitly defined by the parent-child relationships 
in the input taxonomy. 
They capture that a tag in one category implies other tags in the same category.
For instance, the taxonomy may define that the \TAG{adware} class tag is a 
child of the \TAG{grayware} class tag.
Thus, a sample tagged as \TAG{adware} can also be tagged as \TAG{grayware}. 
Inter-category rules capture that a tag from one category 
implies other tags from a different category. 
They make explicit otherwise implicit tag relationships, 
e.g., to add the \TAG{filemodify} behavior to the \TAG{virus} class tag,
as a virus by definition modifies infected files, or
to add the \TAG{spam} behavior to a bot family known to send spam
such as \TAG{conficker}. 
Both inter-category and intra-category expansion rules can be 
automatically identified by the update module. 
In addition, an analyst can provide \tool with its own expansion rules 
based on domain knowledge.

\begin{figure*}[t]
  \small
  \centering
            
\includegraphics[]{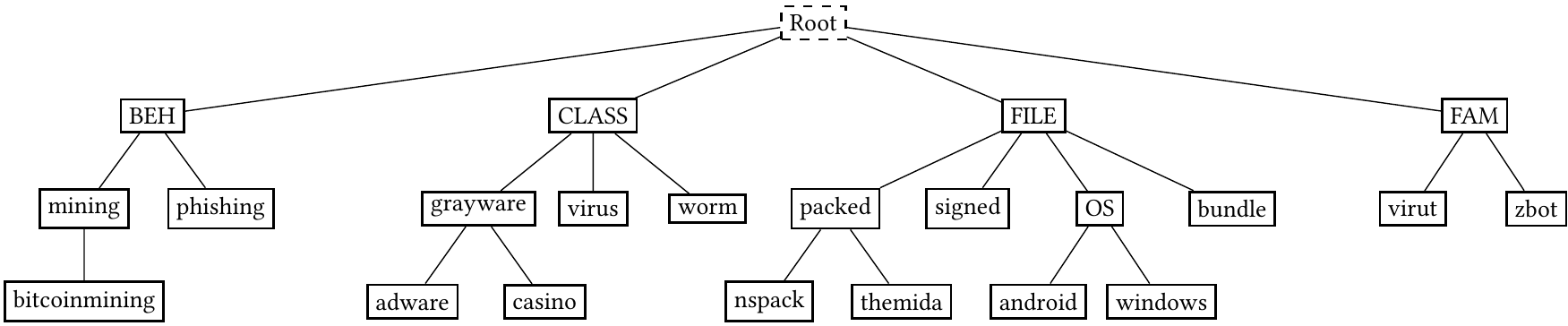}    \caption{A simple taxonomy with all four categories in the default 
            taxonomy, but only a small subset of its tags.}
  \label{fig:taxonomy}
\end{figure*}

\subsection{Taxonomy}
\label{sec:taxonomy}

\Tool takes as input a taxonomy that defines 
parent-child relationships between the tags used by the tagging rules.
\Tool ships with a default taxonomy so that it can be used out-of-the-box. 
The taxonomy supported by \tool is structured as a tree. 
The first level underneath the root comprises categories. 
Each category forms its own subtree where nodes are tags and edges 
capture parent-child relationships between tags. 
Figure~\ref{fig:taxonomy} shows a simplified taxonomy with only a handful 
of tags under each of the four default categories: 
behavior (BEH), class (CLASS), file properties (FILE), and family (FAM).
Some categories such as FAM have no intermediate (i.e., non-leaf) tags, 
they simply contain a set of tags with no structure. 
Other categories contain intermediate tags that capture 
parent-child relationships. 
For example, the CLASS category contains intermediate tag \TAG{grayware} 
indicating that \TAG{adware} and \TAG{casino} 
are classes of grayware.
Some intermediate tags only provide structure and 
are not useful for indexing. 
These appear capitalized, e.g., \TAG{OS}.

By design, \tool assumes that the input taxonomy is incomplete 
and will not contain all malware-related concepts, 
not even all concepts that may appear in AV labels. 
We believe that building a closed malware taxonomy 
that contains all malware-related concepts is a futile effort, 
as it is nearly impossible to be complete and 
malware continuously evolves, requiring constant updates to 
any taxonomy. 
Instead, a key property of \tool is that it supports 
previously unknown concepts, not yet in the tagging rules and taxonomy. 
When an \textit{unknown token}, i.e., a token without a tagging rule, 
appears in an AV label, \tool will keep it and potentially
include it in the output, marking it as having an unknown (UNK) 
pseudo-category. 
Another key property of \tool is that the taxonomy is 
provided as a separate input, so that it can be easily modified.
Thus, we say that \tool uses an {\em open taxonomy} 
that while incomplete, can be refined using the update module 
or in a collaborative manner.

A tag can be represented by its name or by its path in the taxonomy up to 
the category it belongs to.
For example, the \TAG{adware} tag can also be represented as
\TAG{CLASS:grayware:adware}.

\begin{table}[t]
\centering
\small
\caption{Summary of the default taxonomy.}
\begin{tabular}{|l|r|r|p{4.5cm}|}
\hline
  \textbf{Category} &
  \textbf{Tags} &
  \textbf{Leaf} &
  \textbf{Intermediate} \\
  \hline
BEH  & 44 & 38 & \TAG{browsermodify}, \TAG{infosteal}, \TAG{killproc}, \TAG{mining}, \TAG{sms} \\
CLASS & 32 & 23 & \TAG{adware}, \TAG{bot}, \TAG{dialer}, \TAG{grayware}, \TAG{hoax}, \TAG{miner}, \TAG{tool}, \TAG{virus}, \TAG{worm}\\
FAM  & 894 & 894 & - \\
FILE & 95 & 88 & \TAG{exploit}, \TAG{FILETYPE}, \TAG{installer}, \TAG{OS}, \TAG{packed}, \TAG{PROGLANG}, \TAG{signed} \\
\hline
\end{tabular}
\label{tab:categories}
\end{table}

\paragraph{Default taxonomy.}
\Tool provides a default taxonomy,
which organizes concepts that 
commonly appear in AV labels. 
Since the default taxonomy is included in the open source 
release of \tool, 
users can share their updates collaboratively. 
The default taxonomy is summarized in Table~\ref{tab:categories}. 
It comprises of over 1,000 tags split into the four categories 
in Figure~\ref{fig:taxonomy}, which we detail next.

\begin{itemize}

\item {\bf Behavior (BEH).} 
Captures how the malware behaves, 
e.g., \TAG{infosteal}, \TAG{sendssms}, \TAG{spam}, \TAG{mining}.
Behaviors manifest during a sample's execution. 
However, once encoded in an AV label, they can be extracted by \tool 
without the need to execute the sample.

\item {\bf Class (CLASS).}
Malware classes are widely used to capture malware characteristics such 
as specific behaviors or distribution methods. 
Common malware classes are 
\TAG{worm}, \TAG{virus}, \TAG{ransomware}, and \TAG{downloader}. 
A malware family can belong to multiple malware classes. 
For example, \TAG{wannacry} was both \TAG{ransomware} and \TAG{worm}. 
A problematic class is \token{trojan}.
Originally, this class captured a distribution method, 
namely that the sample fooled the user by claiming fake functionality. 
However, nowadays \token{trojan} is the default class used by AV vendors 
for samples without a more specific class. 
Thus, we believe it currently holds little meaning and should be considered 
generic. 
For this reason, \tool by default considers \token{trojan} a generic token,
but the analyst can easily modify this behavior. 

\item {\bf File properties (FILE).}
Comprises of static file properties including 
the file type (e.g., \TAG{pdf}, \TAG{flash}, \TAG{msword}), 
the operating system under which the sample executes
(e.g., \TAG{android}, \TAG{linux}, \TAG{windows}),
the packer used for obfuscation 
(e.g., \TAG{pecompact}, \TAG{themida}, \TAG{vmprotect}), and
the programming language used to code the sample 
(e.g., \TAG{autoit}, \TAG{delphi}, \TAG{java}). 
 
\item {\bf Family (FAM).}
The malware family of the sample. 
The default taxonomy does not include parent-child relationships between 
malware families, i.e., no intermediate family tags. 
Family relationships could be added to the taxonomy using 
malware lineage approaches~\cite{beagle,iline}.

\end{itemize}

The starting point to build the default taxonomy was the file with 
generic tokens in the \avclass repository. 
In \avclass, generics were common tokens in AV labels 
that do not provide family information.
That differs from \tool where generics are tokens that do not provide 
any useful information, family or other.
Our insight was that the \avclass generics contained 
much useful non-family information such as 
malware classes, behaviors, and file properties. 
To build the seed taxonomy,
we examined \avclass generics, manually classifying them into the 
four tag categories above, or as generics that provide no useful information.
After that initial manual effort, 
we refined the seed taxonomy 
by running the update module on different datasets, incorporating 
the newly found tags and relationships, 
as well as adjusting any conflicts it identifies.
Table~\ref{tab:categories} summarizes the current status of the 
default taxonomy.

\subsection{Update Module Overview}
\label{sec:update}

Malware is an ever-evolving ecosystem where new concepts keep appearing.
Over time, known families exhibit new behaviors 
and file properties (e.g., novel obfuscations);
new families are introduced with their corresponding aliases; 
and novel malware classes are occasionally created. 
Furthermore, new relations among known concepts are learned by the 
community increasing our knowledge base 
such as which families belong to a certain class or 
exhibit a specific behavior.
Keeping \tool up-to-date with this natural evolution is a challenge. 
It requires to constantly evolve 
the taxonomy, tagging rules, and expansion rules 
with new concepts and previously unknown relations.
Those new concepts will appear in the output 
of \tool as unknown tokens, not present in the taxonomy and tagging rules.
Manually categorizing those unknown tokens
does not scale to the huge numbers of new samples a security vendor 
may observe each day. 
We need automatic approaches to keep labeling and tagging tools up-to-date.

To this end, \tool provides an update module 
to automatically update the input taxonomy, tagging rules, 
and expansion rules with new concepts and relations.
The update module first identifies co-occurrence relations
of tokens in AV labels. 
Co-occurrence in AV labels was introduced in VAMO~\cite{vamo}
and later used in AVClass~\cite{avclass} and Euphony~\cite{euphony} 
to identify family aliases. 
But, the update module in \tool takes this concept a step further by 
introducing a novel recursive process that first identifies 
relations between unknown tokens and tags and then uses a 
set of learning rules to classify those relations and propose 
updates to the taxonomy, tagging rules, and expansion rules.

\begin{figure}[!t]
\includegraphics[width=\columnwidth]{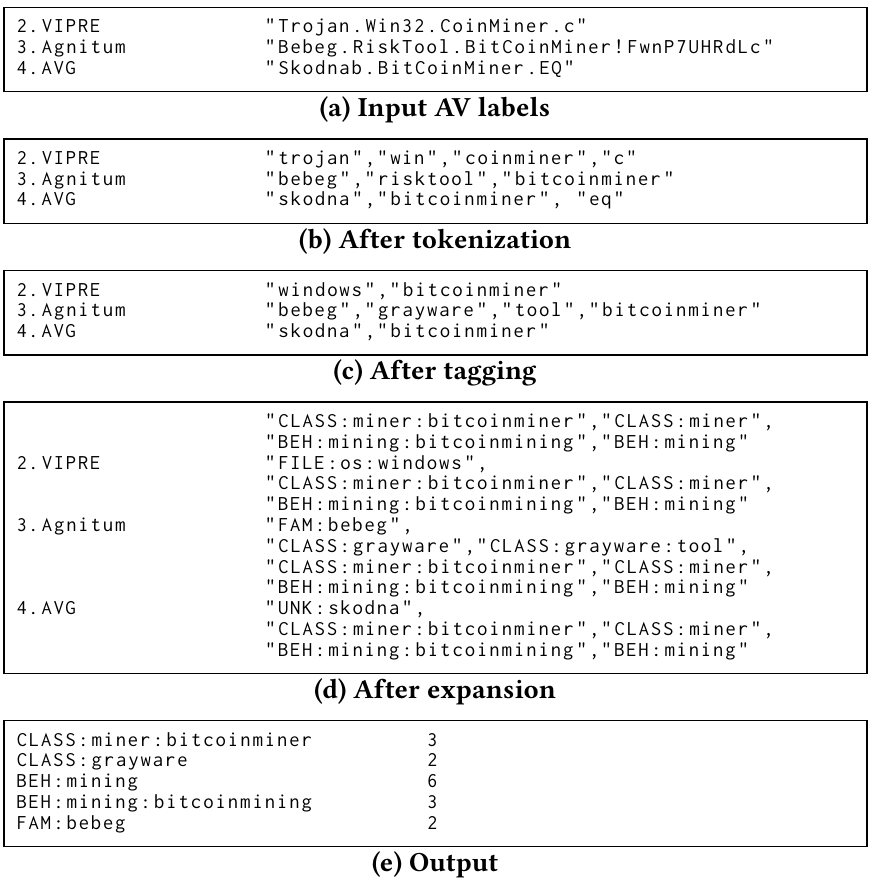}
\caption{Running example.}
\label{fig:running}
\end{figure}

\section{Labeling Module}
\label{sec:approach}

The labeling module takes as input the AV labels assigned by
multiple AV engines to the same set of samples,
an optional list of AV engines whose labels to use,
a set of tagging rules,
an optional set of expansion rules, and
a taxonomy.
For each input sample, it outputs a set of tags
ranked by the number of AV engines
including the tag's concept in their label.
\tool comprises of three steps:
tokenization, tagging, and expansion. 
To illustrate them we use the running example in Figure~\ref{fig:running}.
The inputs are the labels assigned by four AV engines to the same sample, 
shown in Figure~\ref{fig:running}a.

\paragraph{Tokenization.}
Takes as input an AV label and outputs the list of tokens 
the label contains. 
The tokenization in \tool is almost identical to that of \avclass 
and we refer the reader to the original paper for details~\cite{avclass}. 
The only significant difference is that \tool does not filter out 
short tokens (less than four characters) during tokenization, 
but rather after tagging. 
This enables \avclass to extract tags from short tokens that correspond 
to well-known concepts, e.g., \token{irc}, \token{bot}.
Figure~\ref{fig:running}b shows the tokens obtained in our 
running example.

\paragraph{Tagging.}
Takes as input the tokens obtained from the tokenization and 
the input tagging rules. 
For each input token, if a tagging rule exists for the token, 
it applies it to obtain a list of tags. 
If the token is generic, it is removed.
If no tagging rule exists for the token, it is kept. 
It outputs a list of identified tags and any remaining unknown tokens.
Figure~\ref{fig:running}c shows the tagging output. 
Some tokens have been replaced by their tags, 
e.g., token \token{win} by tag \TAG{windows} and 
token \token{risktool} by tags \TAG{grayware} and \TAG{tool}.
Other tokens have been dropped 
as generic (e.g., \token{trojan}) or because they are 
short (e.g., \token{eq}).

\paragraph{Expansion.}
Takes as input the file with expansion rules, 
the taxonomy, and the tags output by the tagging. 
For each tag, if an expansion rule exists for it, 
it applies the rule to obtain a larger list of target tags that 
replaces the tag.
Unknown tokens are not affected by the expansion. 
The expansion first applies the inter-category expansion rules 
provided as input to \tool. 
Next, it applies the implicit intra-category expansion rules 
due to the parent-child tag relationships in the taxonomy.
For example, in our running example tag \TAG{bitcoinminer} 
implies tag \TAG{CLASS:miner}.

\paragraph{Output.}
For each tag and unknown token, \tool counts the number of AV labels 
where it appears. 
This count can be interpreted as a confidence score.
Tags and unknown tokens that appear in the label of at most one AV engine 
are removed.
This filters random unknown tokens that earlier steps may have missed, 
as the likelihood that those appear in labels from multiple AV engines is low, 
as well as very low confidence tags.

The output of \tool is the list of tags and unknown tokens along with their 
confidence score.
Figure~\ref{fig:running}e shows the output of our running example 
where unknown token \token{skodna}, 
as well as tag \TAG{FILE:os:windows}, 
have been removed because they only appeared in one label.
\tool also provides a compatibility mode with \avclass 
to output the most likely family for each sample, 
which corresponds to the highest ranked family tag or unknown token.
i.e., \TAG{FAM:bebeg} in our running example.

\section{Update Module}
\label{sec:aliases}

The update process comprises of two steps. 
When labeling a dataset, 
\tool outputs co-occurrence statistics
between tags and unknown tokens. 
The larger the dataset, 
the higher confidence in the identified co-occurrence statistics. 
The update module takes as input the co-occurrence statistics, 
the taxonomy, and the tagging and expansion rules. 
It performs two substeps: 
identifying strong relations and 
converting strong relations into updates to the input files. 
The process of generating co-occurrence statistics and 
identifying strong relations is similar to the one used by 
\avclass to detect alias relations~\cite{avclass}. 
The novel part of the update module is the recursive process and 
update rules used to automatically generate updates 
to the taxonomy, tagging rules, and expansion rules.

\paragraph{Co-occurrence statistics.}
The labeler obtains the co-occurrence statistics after tagging and 
before expansion. 
For simplicity, in this section we call tags to both tags and unknown tokens.
We call {\em relation} to each pair of tags 
that appear in labels for the same sample and its co-occurrence statistics. 
For each relation, the labeler outputs seven values: 
the tags $t_i$, $t_j$, 
the number of samples each tag appears in the dataset being labeled 
$|t_i|$, $|t_j|$, 
the number of samples where both tags appear $|(t_i, t_j)|$, and
the fraction of times that both tags appear in the same samples 
$rel(t_i,t_j) = \frac{|(t_i, t_j)|}{|t_i|}$ and 
$rel(t_j,t_i) = \frac{|(t_i, t_j)|}{|t_j|}$. 
The two tags are sorted so that tag $t_i$ is the one that occurs less often, 
i.e., $|t_i| \le |t_j|$, which means that $rel(t_i,t_j) \ge rel(t_j,t_i)$.

\paragraph{Identifying strong relations.}
Given the set of relations output by the labeler, 
the update module first filters out weak relations.
A relation is {\em strong} 
if both tags have been seen in enough number of samples and
appear in the same samples frequently enough.
The first condition keeps only relations where 
$min(|t_i|$,$|t_j|) \ge n$ where $n$ is the minimum number of samples where 
a tag should have been observed. 
The second condition keeps only relations where $rel(t_i,t_j) \ge T$. 
Threshold $T$ controls the minimum joint frequency to determine a 
relation is strong.
For strong relations, we say that $t_i$ {\em implies} $t_j$, $t_i \Rightarrow t_j$, 
but $t_j$ may not imply $t_i$. 
For example, if family tag \TAG{virut} appears in 1M samples and 
class tag \TAG{virus} in 7M samples, and in every sample \TAG{virut} appears 
\TAG{virus} also appears ($rel(virut, virus) = 1.0$), 
then \TAG{virut} \IMPLIES \TAG{virus}, but 
\TAG{virus} $\not \Rightarrow$ $virut$ as 
there are 6M instances were \TAG{virus} is observed without 
\TAG{virut}, likely corresponding to other virus families
($rel(virus, virut) = 0.14$).
If both $rel(t_i,t_j) \ge T$ and $rel(t_j,t_i) \ge T$ 
we say the tags are {\em equivalent}, $t_i \Leftrightarrow t_j$.
Parameters $n$ and $T$ were empirically selected in \avclass 
and we use their suggested default values of $n=20$ and $T=0.94$.

Weak relations and relations including a OS tag 
are removed. 
The latter avoids that an expansion rule is created 
for each family towards its platform, 
e.g., \TAG{virut} \IMPLIES \TAG{windows}, 
\TAG{droidkungfu} \IMPLIES \TAG{android}. 

\begin{table}[t]
\caption{Update module rules.} 
\small
\centering
\begin{tabular}{|l|l|l|}
\hline
{\bf Cat($t_i$)} & {\bf Cat($t_j$)} & {\bf Rule} \\
\hline
UNK & FAM & tagging($t_i$, $t_j$, FAM) \\
UNK & CLASS & taxonomy($t_i$, FAM:$t_i$) \\ UNK & BEH & taxonomy($t_i$, FAM:$t_i$) \\ UNK & FILE & taxonomy($t_i$, path($t_j$):$t_i$)  \\
UNK & UNK & taxonomy($t_i$, FAM); taxonomy($t_j$, FAM) \\ FAM & UNK & tagging($t_i$, $t_j$, FAM) \\
FILE & UNK & tagging($t_i$, $t_j$, prefix($t_i$)) \\
FAM & FAM & tagging($t_i$, $t_j$, prefix($t_j$)) \\
\hline
\hline
FAM & FILE & expansion($t_i$, $t_j$) \\
FAM & BEH & expansion($t_i$, $t_j$) \\
FAM & CLASS & expansion($t_i$, $t_j$) \\
CLASS & FILE & expansion($t_i$, $t_j$) \\
CLASS & BEH & expansion($t_i$, $t_j$) \\
\hline
\end{tabular}
\label{tab:alias_rules}
\end{table}

\paragraph{Updates to taxonomy and tagging.}
The update module performs a recursive process 
where each iteration examines the set of remaining relations. 
The process starts with the identified strong relations. 
At each iteration, every remaining relation is checked against a set of 
rules to identify updates to the current 
taxonomy, tagging, and expansion rules. 
Processed relations are removed, the rest are kept.
Once all relations are examined, the process runs 
into a new iteration with the remaining relations.
Recursion ends when one iteration does not produce any updates or 
no relations remain.

Each relation is first checked to see if it is already known. 
A relation is known if it is already 
captured in the current taxonomy, tagging rules, or expansion rules.
For example, \TAG{adware} \IMPLIES \TAG{grayware} is implicit in the default taxonomy.  
This check happens before each relation is processed because 
the taxonomy, tagging, and expansion rules 
change as relations are processed and a relation that was not known before 
may become known once other relations have been processed. 
Known relations are removed.
If not known, and the tags are equivalent, 
a tagging rule is added from $t_i$ to $t_j$
since $t_i$ is the least common of the two tags.
If not an equivalence, the relation is processed according to the rules in 
the top block of Table~\ref{tab:alias_rules}, 
which are indexed by the categories of the two tags. 
UNK is a pseudo-category for unknown tags not in the taxonomy 
(i.e., unknown tokens).
All rules, but the last one, handle relations where at least one tag 
is unknown. 
These rules create a tagging rule between the two tags or 
add the unknown tag to the taxonomy with a path prefix indicated 
by the rule.
For example, the top rule captures that an unknown tag implies a 
family tag, which creates a tagging rule capturing that the unknown tag $t_i$ 
is an alias for the family tag $t_j$. 
Adding a tagging rule $t_i \Rightarrow t_j$ forces 
$t_i$ to be removed from the taxonomy (if present) and 
$t_j$ to be added (if it did not exist). 
The last rule in the block creates a tagging rule for a relation 
between family tags indicating that $t_i$ is an alias for $t_j$ and 
thus $t_i$ does not need to be in the taxonomy.

\paragraph{Updates to expansions.}
Since expansions happen between two tags in the taxonomy 
it is more efficient to identify them once all rules for 
unknown tags have been applied and the taxonomy and tagging rules are 
stable.
Once the recursion ends, all remaining rules are examined using the 
expansion rules in the bottom block of Table~\ref{tab:alias_rules}. 
The expansion rules capture properties of a family such as its class, 
file properties or behaviors, as well as a class having specific 
behaviors or file properties (e.g., using a exploit).

The five categories (including UNK), create 25 distinct category pairs. 
However, only 13 pairs are considered in Table~\ref{tab:alias_rules}. 
There exist 12 category pairs without a rule. 
Our evaluation will show that those pairs constitute less than 1\% of 
relations and when they happen they indicate some collision that we 
believe is best resolved manually by an analyst. 

Once the process finishes the update module outputs the updated 
taxonomy, tagging and expansion rules. 
Most updates will be additional tags in the taxonomy and new tagging and 
expansion rules, but it is possible that some original 
taxonomy entries and tagging rules have been modified or removed.

\section{Evaluation}
\label{sec:evaluation}

This section evaluates \tool. 
First, Section~\ref{sec:datasets} presents the datasets used. 
Then, Section~\ref{sec:eval_tagging} details the tagging results
and Section~\ref{sec:eval_alias} demonstrates the update module. 
Section~\ref{sec:eval_search} illustrates the benefits of \tool. 
Finally, Section~\ref{sec:eval_famlabels} compares \tool with prior 
family tagging tools \avclass and \euphony.

\begin{table}[t]
\caption{Datasets used in evaluation.
}
\small
\centering
\begin{tabular}{|l|l|c|r|c|l|}
\hline
  \textbf{Dataset}
  &\textbf{Plat.}
  &\textbf{GT}
  &\textbf{Samples}
  &\textbf{Bin.}
  &\textbf{Collection} \\
\hline
\superset                             & Mix & \N & 42,533,619 & \N & 08/2008 - 05/2019\\
Lever et al.~\cite{candia}            & Win & \N & 37,817,328 & \N & 01/2011 - 08/2015\\
Andropup              & And & \N & 3,145,283 & \N  & 05/2019 - 09/2019\\
Miller et al.~\cite{miller16reviewer} & Win & \N & 1,079,783 & \N  & 01/2012 - 06/2014\\
Andrubis~\cite{andrubis}              & And & \N & 422,826   & \N  & 06/2012 - 06/2014 \\
Malsign~\cite{malsign}                & Win & \Y & 142,500   & \N  & 06/2012 - 02/2015 \\
AMD~\cite{amd}                        & And & \Y & 24,553    & \Y  & 11/2010 - 03/2016\\
Malicia~\cite{malicia}                & Win & \Y & 9,908     & \Y  & 03/2012 - 02/2013\\
Drebin~\cite{drebin}                  & And & \Y & 5,560     & \Y  & 08/2010 - 10/2012\\
Malheur~\cite{malheur}                & Win & \Y & 3,131     & \N  & 08/2009 - 08/2009\\ MalGenome~\cite{malgenome}            & And & \Y & 1,260     & \Y  & 08/2008 - 10/2010\\
\hline
\end{tabular}
\label{tab:datasets}
\end{table}

\subsection{Datasets}
\label{sec:datasets}

We evaluate \tool using the \numdatasets datasets in  
Table~\ref{tab:datasets}.
For each dataset, the table shows the target architecture of the samples 
(Windows, Android, or both), 
whether the samples are labeled with a family name, 
the number of samples, 
whether binaries are available 
(otherwise we only have their hashes), and 
the collection period.
Datasets come from prior works, 
save for Superset that is the union of all other datasets.
Some datasets overlap,
e.g., Drebin~\cite{drebin} is a superset of MalGenome~\cite{malgenome}.
We do not remove duplicates to make it easy to map results 
to publicly available datasets.

We had VT reports for most samples and only collect a few missing ones. 
We use the available reports
since VT's rate limits make it infeasible for us to collect 
the latest reports.

\begin{table}[!t]
  \caption{Percentage of samples for which a tag could be extracted, as well as percentage of tagged samples with a tag from each category.}
  \label{tab:tagging}
  \small
  \centering
  \begin{tabular}{|r|r||r|r|r|r|r|r|}
    \cline{2-3}
    \multicolumn{1}{c|}{} & \multicolumn{2}{c|}{{\bf Tagged}} & \multicolumn{5}{c}{} \\
    \hline
    \textbf{Dataset}
      &\textbf{All}
      &\textbf{VT$\ge$4}
      &\textbf{FILE}
      &\textbf{CLA}
      &\textbf{BEH}
      &\textbf{FAM}
      &\textbf{UNK}\\
    \hline
    \superset      & 97\% & 100\% & 99\% & 94\% & 75\% & 83\% & 13\% \\
    Lever          & 98\% & 100\% & 99\% & 94\% & 78\% & 83\% & 13\% \\
    Andropup       & 89\% & 99\% & 98\% & 95\% & 28\% & 74\% & 16\% \\
    Miller         & 95\% & 99\% & 100\% & 91\% & 82\% & 82\% & 20\% \\
    Andrubis      & 100\% & 100\% & 100\% & 89\% & 69\% & 98\% & 4\% \\
    Malsign        & 99\% & 99\% & 97\% & 95\% & 73\% & 96\% & 5\% \\
    AMD            & 100\% & 100\% & 100\% & 99\% & 31\% & 98\% & 14\%\\
    Malicia       & 100\% & 100\% & 100\% & 89\% & 85\% & 97\% & 12\%\\
    Drebin        & 100\% & 100\% & 100\% & 92\% & 71\% & 98\% & 5\% \\
    Malheur        & 100\% & 100\% & 100\% & 93\% & 83\% & 66\% & 7\% \\
    MalGen.        & 100\% & 100\% & 100\% & 96\% & 82\% & 98\% & 2\% \\
    \hline
  \end{tabular}
\end{table}

\begin{table*}[t]
  \caption{Top 10 tags for each category ranked by the number of samples assigned the tag in the Superset dataset.}

  \small
  \centering
  \begin{tabular}{|r||rr||rr||rr||rr||rr|}
    \cline{1-11}
      \multicolumn{1}{|c||}{\bf Rank}
        & \multicolumn{2}{c||}{\bf FILE}
        & \multicolumn{2}{c||}{\bf CLASS}
        & \multicolumn{2}{c||}{\bf BEH}
        & \multicolumn{2}{c||}{\bf FAM}
        & \multicolumn{2}{c|}{\bf UNK}\\
      \hline
1 & windows & 61\% & grayware & 46\% & execdownload & 27\% & vobfus & 10\% & fraudpack & 1\%\\
2 & packed & 20\% & downloader & 20\% & filemodify & 21\% & loadmoney & 5\% & atraps & 1\%\\
3 & android & 5\% & adware & 18\% & selfpropagate & 12\% & virut & 5\% & hiddapp & 1\%\\
4 & bundle & 3\% & virus & 15\% & autorun & 10\% & softpulse & 4\% & hiddenads & 1\%\\
5 & installer & 2\% & worm & 9\% & inject & 6\% & installerex & 3\% & packer & 1\%\\
6 & small & 2\% & backdoor & 4\% & server & 6\% & domaiq & 3\% & llac & 1\%\\
7 & nsis & 1\% & multiplug & 4\% & alertuser & 2\% & firseria & 3\% & trymedia & 1\%\\
8 & msil & 1\% & rogueware & 2\% & killproc & 2\% & zbot & 3\% & refroso & 1\%\\
9 & autoit & 1\% & tool & 1\% & sms & 2\% & sality & 3\% & bifrost & 1\%\\
10 & html & 1\% & clicker & 1\% & ddos & 2\% & virlock & 2\% & comame & 1\%\\
  \hline
  \end{tabular}
  \label{tab:accuracy}
\end{table*}

\subsection{Tagging Evaluation}
\label{sec:eval_tagging}

Table~\ref{tab:tagging} summarizes the tagging coverage of \tool, 
i.e., the fraction of samples for which it can extract a tag. 
These results are obtained using the tagging rules, 
expansion rules, and taxonomy output by the update module 
after it identifies previously unknown relations.
In Section~\ref{sec:eval_alias} we analyze the tagging improvement 
before and after applying the update module.
The table first shows the fraction of all samples, 
and those flagged by at least four AV engines,
for which at least one tag was obtained. 
We use the threshold of four AV detections to remove potentially 
benign samples.
This threshold has been used in prior works~\cite{malsign} 
and a recent work shows that threshold values between two and 14 are good
for stability and for balancing precision and recall~\cite{zhu20labels}.
For each category, the table then shows the fraction of tagged samples 
with at least four detections with a tag of that category. 
Column UNK corresponds to samples with at least one output unknown token 
not in the taxonomy.

The results show that \tool can extract at least one tag for 89\%--100\% of 
the samples, depending on the dataset. 
Thus, it is possible to index the majority of samples. 
The files for which no tags can be extracted largely correspond to those 
with very few detections, which is higher for Andropup as it is the most 
recent dataset. 
When considering sample flagged by at least four AV engines the 
fraction of tagged samples is at least 99\%.
As shown by the Superset, 
the most common tags are 
file properties (99\% of samples), followed by 
malware classes (94\%),
known families (83\%), and 
behaviors (75\%).
It is important to note that the FAM column 
considers only samples tagged with families that appear in the taxonomy. 
However, unknown tags most often correspond to new families 
that have not yet been added to the taxonomy and should be 
considered for final family tagging results. 
Section~\ref{sec:eval_famlabels} compares the family tagging results
of \tool to prior tools like \avclass and \euphony.

\paragraph{Most popular tags.}
Table~\ref{tab:accuracy} shows the top 10 tags in the Superset 
for each category, 
ranked by the number of samples assigned the tag. 
The most popular tag is \TAG{FILE:OS:windows} (61\% of samples), 
followed by \TAG{CLASS:grayware} (46\%), 
\TAG{BEH:execdownload} (27\%), 
\TAG{BEH:filemodify} (21\%), 
\TAG{FILE:packed} (20\%), and 
\TAG{CLASS:downloader} (20\%).
Our taxonomy contains 32--95 tags in each of the non-family categories, and 
894 families. 
Of the tags in the taxonomy 94\%--100\%, depending on the category, 
appear in the Superset. 
Thus, the distribution of tags per category contains a large 
number of tags with at most 1\% coverage, 
each identifying up to tens of thousands of samples. 
Thus, \tool extracts a wide variety of tags that can be used by the 
analyst to search for samples according to class, family, file properties,  
and behaviors.

The top FILE tags include the platform (\TAG{windows}, \TAG{android}), 
whether the sample is \TAG{packed}, 
if it is a \TAG{bundle} that contains other executables, 
the programming language (\TAG{autoit}, \TAG{msil} for C\#), 
if it is an \TAG{installer} or has been generated by a particular 
installer (\TAG{nsis}), and
the size of the file (\TAG{small}). 
Intermediate tags rank high because 
they accumulate popularity of their children 
through the expansion on taxonomy relationships. 
For example, \TAG{packed} accumulates the influence of all 
packer tags in the taxonomy 
(e.g., \TAG{asroot}, \TAG{upack}, \TAG{themida}).
And, \TAG{installer} the influence of \TAG{nsis}, 
as well as other installer-generating software 
(e.g., \TAG{wiseinstaller}, \TAG{installmate}).

There are four CLASS tags appearing in more than 10\% of the samples. 
They capture the popularity of potentially unwanted programs, 
downloaders, monetizing through advertisements, and 
viruses.
\TAG{CLASS:grayware:adware:multiplug} are 
\TAG{adware} that install browser plugins (e.g., extensions, toolbars) to 
modify the user's Web surfing.
And, \TAG{CLASS:grayware:tool} are tools not necessarily 
malicious, but often abused, such as those used for remote administration.
Note that \TAG{trojan} is considered generic, 
otherwise it would be assigned to 86\% of the samples. 

The top three behavior tags correspond to expansions from class tags:
\TAG{downloader} \IMPLIES \TAG{execdownload}, 
\TAG{virus} \IMPLIES \TAG{filemodify}, 
\TAG{worm} \IMPLIES \TAG{selfpropagate}. 
Note that the behavior associated to a class 
gets a higher tagging ratio than its class, 
indicating additional samples without the corresponding class tag. 
For example, \TAG{execdownload} has an additional 7\% samples 
without the \TAG{downloader} class tag and 
\TAG{filemodify} an additional 6\% over \TAG{virus}.
Other popular behaviors are \TAG{autorun}, which captures samples that 
modify the {\tt autorun.inf} Windows file to automatically execute; 
those related to obfuscation such as injecting 
code in a benign process and killing a process, 
typically of a security tool;
opening a server, 
alerting the user, 
sending SMS messages; and
launching denial of service attacks. 

Top families have lower prevalence than top 
classes, file properties, and behaviors.
The most prevalent family is \TAG{vobfus} (10\%).
Out of the top families, half correspond to \TAG{grayware} families
(\TAG{loadmoney}, \TAG{softpulse}, \TAG{installererex}, 
\TAG{domaiq}, \TAG{firseria}) and the rest are malware. 
Top unknown tokens have much lower prevalence, at most 1\%. 
They mostly correspond to families not yet in the taxonomy, 
for which no strong relation has yet been observed to another tag.

\begin{table*}[t]
\caption{Update module evaluation results 
with $n = 20$ and $T = 0.94$.}
\small
\centering
\begin{tabular}{|l|r|r|r|r|r|r|r|r|r|}
\cline{2-10}
\multicolumn{1}{c|}{}
& \multicolumn{3}{c|}{\bf Relations}
& \multicolumn{2}{c|}{\bf Taxonomy Entries}
& \multicolumn{2}{c|}{\bf Tagging Rules}
& \multicolumn{2}{c|}{\bf Expansion Rules} \\
\hline
{\bf Dataset} & {\bf All} & {\bf Strong} & {\bf Out} &
{\bf Added} & {\bf Rem.} & 
{\bf Added} & {\bf Rem.} & 
{\bf Added} & {\bf Rem.} \\
\hline
Andropup & 30,107 & 968 & 3 & 486 & 2 & 216 & 10 & 461 & 0 \\
\hline
\end{tabular}
\label{tab:alias_summary}
\end{table*}

\subsection{Update Module}
\label{sec:eval_alias}

We illustrate the usage of the update module with the Andropup dataset. 
When we first obtained this dataset, 
we run \tool's labeler observing that 65\% of samples contained 
an unknown tag.
To reduce this number we run the update module on the current taxonomy 
and without any expansion rules. 
The updated taxonomy, tagging rules, and expansion rules
reduced the samples with an unknown tag 
from 65\% to the 16\% shown in Table~\ref{tab:tagging}.

Table~\ref{tab:alias_summary} summarizes the update module results.
Since we only have the same family ground truth used in 
\avclass, we use the default values suggested in that work 
of $n = 20$ and $T = 0.94$~\cite{avclass}. 
The co-occurrence statistics output by the labeler contain 30,107 relations, 
of which 968 are strong. 
Those 968 relations belong to 11 category pairs, the most popular being 
UNK \IMPLIES CLASS (61\%), 
UNK \IMPLIES UNK (17\%), 
UNK \IMPLIES BEH (9\%), 
UNK \IMPLIES FAM (7\%), 
FAM \IMPLIES CLASS (2\%), and
UNK \IMPLIES FILE (2\%).
The other five category pairs had at most 0.3\% relations each.
Overall, 96\% of the strong relations involve an unknown token.

From those 968 strong relations, the update module automatically identified
486 new taxonomy entries, 216 new tagging rules, and 461 expansion rules.
Of the new taxonomy entries, 97\% correspond to new families.
The other 3\% are file properties, mostly Android exploits,
e.g., \TAG{FILE:exploit:asroot}, \TAG{FILE:exploit:exploid}, and 
\TAG{FILE:exploit:gingerbreak}.
The new expansion rules link families to their class or behavior. 
The most popular destination tags are:
\TAG{grayware} (46\%), \TAG{adware} (26\%), \TAG{infosteal} (8\%), and
\TAG{downloader} (6\%).

At the end of the update module processing, there remain only three 
strong relations (0.3\%) for which no updates are generated because they 
lack a processing rule. 
For these relations the analyst would have to manually decide how 
to handle them.
The first is \TAG{FILE:patch} \IMPLIES \TAG{CLASS:grayware}. 
This is an example of an \textit{homonym}, 
i.e., a token with two possible meanings. 
In our seed taxonomy, obtained from the \avclass generics,
we manually (incorrectly) classified \TAG{patch} as a file type. 
But, this relation automatically extracted by the update module allowed 
us to understand that the tag is used for modding apps 
(e.g., Lucky Patcher\footnote{https://www.luckypatchers.com/}) 
that patch other apps for unlimited access, no ads, etc.
This allowed us to correct our error making \token{patch}
an alias for \TAG{BEH:filemodify}.
Another homonym we incorrectly classified in our seed 
taxonomy is \token{fakedoc}, which we though meant a fake document, 
but the update module identified that 
\TAG{batterydoctor} \IMPLIES \TAG{FILE:fakedoc},
indicating \TAG{fakedoc} is instead a rogueware family (\TAG{fakedoctor}). 
The other two unhandled relations are due to over-fitting of a particular 
family. 
Relation \TAG{FILE:proglang:powershell} \IMPLIES \TAG{CLASS:keylogger}
is due to a keylogging family that uses Powershell scripts and 
relation \TAG{BEH:inject} \IMPLIES \TAG{CLASS:downloader} 
to a downloader family that injects in other processes. 
Since they are not general enough, we ignore both relations.

\paragraph{Update quality.}
To evaluate the quality of the generated updates, 
we manually examine the resulting taxonomy, tagging, and expansion rules 
looking for errors based on our domain knowledge and Web searches. 
We acknowledge that this process is subjective and may not spot all errors, 
but it is the best ground truth we were able to obtain.
We identify 11 cases were we would have done things differently than the 
update module. 
First, five family tags should likely be classes/subclasses
(\TAG{bankbot}, \TAG{clickfraud}, \TAG{fakeantivirus}, \TAG{locker}, 
\TAG{remoteadmin}).
In addition, \TAG{trojandldr} was marked as a family tag with an expansion 
rule \TAG{trojandldr} \IMPLIES \TAG{downloader}, but making 
the expansion rule a tagging rule would have been cleaner. 
One challenge is distinguishing the name of an exploit (file property) 
from a family that uses an exploit. 
In most cases the update module is correct, but 
\TAG{FILE:exploit:rootmaster} should likely be 
\TAG{FAM:rootmaster} with an expansion rule 
\TAG{rootmaster} \IMPLIES \TAG{exploit}. 
There is also one tagging rule \TAG{cryptominer} \IMPLIES \TAG{coinhive} 
where \TAG{cryptominer} should likely be an alias for \TAG{CLASS:miner}.
One new file property in the taxonomy \TAG{FILE:proglang:java:genericgba}
is likely a generic token.
And, another file property \TAG{FILE:packed:decrypter} 
does not look like any known packer, so we are not sure what it 
exactly represents.
Finally, \TAG{FILE:testvirus:testfile} would be better as a tagging rule
\TAG{testfile} \IMPLIES \TAG{testvirus}.

In summary, out of 1,163 updates to the taxonomy and rules, 
only 11 (0.9\%) required adjustments. 
Additionally, three relations had to be checked, 
allowing us to correct errors with homonyms introduced in the manual 
creation of the seed taxonomy.

\subsection{Beyond Family Labeling}
\label{sec:eval_search}

The main benefits of \tool over prior family labeling tools 
are that it enables tag-based searches beyond families, and 
it builds malware knowledge generalizing beyond individual samples.

\paragraph{Search.}
The tags output by \tool enable advanced searches on a malware repository. 
The 975 tags identified in the Superset, 
can be used, among others, to identify samples that 
are grayware, 
belong to a specific class 
(e.g., ransomware, information stealers), 
send SMS, 
launch DoS attacks,
use a specific packer (e.g., \TAG{themida}, \TAG{asroot}), or 
installer software (e.g., \TAG{nsis}, \TAG{wiseinstaller}) software. 
Those samples can then be used to build classifiers. 

We examine how the extracted tags compare with those already used
by VirusTotal.
The VT documentation mentions 335 tags,
mostly corresponding to file properties and behaviors~\cite{vtTags}.
Those 335 tags do not include family names or classes 
(save for \TAG{worm} and \TAG{emailworm}).
Of those 335 tags, we observe 259 in the reports of samples in the 
Superset.
Of those VT tags, 49 are identical to tags in our default taxonomy 
with the largest types being  
30 packers,
4 programming languages, and
4 file types.
Thus, the tags output by \tool nicely complement 
the ones already in use by VT. 
Since \tool already operates on VT reports, adding \tool's tags to VT 
would be straightforward and would enable new searches not currently possible 
such as those for samples of a particular malware class or family, 
as well as searches for a richer set of file properties and behaviors.
This highlights the benefit of \tool's automatic tag extraction 
to popular malware repositories.  

\paragraph{Knowledge base.}
\Tool taxonomy, tagging rules, and expansion rules form a 
malware knowledge base that capture relations beyond individual samples 
such as which families are ransomware or information stealers.
Compared to existing online malware encyclopedias, 
its contents have been obtained using a well-defined methodology; 
are not specific to one vendor; and
since they are publicly available, 
can be discussed and evolved collaboratively.
Furthermore, \tool's update module allows to refine
the knowledge base over time keeping it up-to-date.

\begin{table*}[!t]
\caption{Family labeling comparison between \tool, \avclass, and \euphony on 
datasets with ground truth. T is in seconds.}
\small
\centering
\resizebox{\textwidth}{!}{
\begin{tabular}{|r|r|r|r|r|r||r|r|r|r|r||r|r|r|r|r||r|r|r|r|r|}
\cline{2-21}
\multicolumn{1}{c|}{}
& \multicolumn{5}{c||}{\bf \tool }
& \multicolumn{5}{c||}{\bf \avclass*}
& \multicolumn{5}{c||}{\bf \avclass}
& \multicolumn{5}{c|}{\bf \euphony} \\
\hline
\textbf{Dataset}
&\textbf{T}
&\textbf{Fam}
&\textbf{Prec}
&\textbf{Rec}
&\textbf{F1}
&\textbf{T}
&\textbf{Fam}
&\textbf{Prec}
&\textbf{Rec}
&\textbf{F1}
&\textbf{T}
&\textbf{Fam}
&\textbf{Prec}
&\textbf{Rec}
&\textbf{F1}
&\textbf{T}
&\textbf{Fam}
&\textbf{Prec}
&\textbf{Rec}
&\textbf{F1}\\
\hline
Malsign   & 205 & 99\% & 91\% & 92\% & 92\% & 139 & 99\% & 91\% & 92\% & 92\% & 139 & 100\% & 90\% & 90\% & 90\% & 2,126 & 97\% & 74\% & 49\% & 59\%\\
AMD       & 44 & 100\% & 92\% & 80\% & 86\% & 50 & 100\% & 92\% & 80\% & 86\% & 50 & 100\% & 92\% & 76\% & 83\% & 331 & 100\% & 76\% & 70\% & 73\%\\
Malicia   & 20 & 100\% & 95\% & 60\% & 74\%  & 14 & 100\% & 95\% & 60\% & 74\% & 14 & 100\% & 95\% & 61\% & 74\% & 593 & 100\% & 86\% & 54\% & 67\% \\
Malheur   & 2 & 100\% & 91\% & 94\% & 92\% & 1 & 100\% & 91\% & 94\% & 92\% & 1 & 100\% & 91\% & 98\% & 94\% & 53 & 100\% & 90\% & 93\% & 91\% \\
Drebin    & 5 & 100\% & 94\% & 96\% & 95\% & 8 & 100\% & 94\% & 96\% & 95\% & 8 & 100\% & 96\% & 89\% & 92\% & 208 & 100\% & 94\% & 85\% & 89\% \\
\cline{1-16}
\hline
\end{tabular}
}
\label{tab:gtEuphony}
\end{table*}

\subsection{Family Labeling Comparison}
\label{sec:eval_famlabels}

This section compares \tool with \avclass and \euphony for family labeling.
The goal is not to evaluate \tool for family labeling, 
as an analyst that only requires family labeling, 
but not the additional malware and threat intelligence \tool provides,
does not need to use \tool, 
as it does not significantly differ from \avclass for that task. 
Instead, the comparison allow us to 
(1) evaluate \avclass versus \euphony
(not done in the original works), 
(2) highlight the importance of the update module, and 
(3) show our vision for using \tool and \avclass.


For the comparison, we use the most recent versions of 
\euphony~and \avclass~at the time of writing.
and the Windows and Android datasets with ground truth. 
While \euphony was only evaluated on labels of Android samples, 
its processing is equally valid for Windows.
For \tool, we use its \avclass compatibility option, 
which selects as the sample's family the highest ranked 
family tag or unknown token.
For \avclass, we use the most recent aliases and generics in its repository. 
We also include in the comparison {\avclass}*, 
a modified version of \avclass that 
takes as input the taxonomy and tagging rules from \tool, 
instead of the alias and generic files in the \avclass repository. 
{\avclass}* extracts from the taxonomy and tagging rules the list of 
aliases and generic tokens \avclass needs. 
The reason to include {\avclass}* is to compare \tool and \avclass 
on the same input knowledge base, 
ignoring any differences due to the knowledge base of \tool 
being more up-to-date. 

Results of the four tools, run on the same server,
are in Table~\ref{tab:gtEuphony}.
For each tool and dataset it shows the runtime in seconds, 
the percentage of samples with a family, and the accuracy 
(precision, recall, F1 score) when the results are compared with the 
available ground truth.
Given that these are old datasets with large average number of detections 
per sample, all tools output a label for almost all samples. 
\Tool and {\avclass}* achieve the highest F1 score on four of the five datasets,
while \avclass ranks first on Malheur.
\euphony's accuracy drops significantly on the larger datasets from 
89\%-91\% on Malheur and Drebin down to 59\%-73\% on Malsign, Malicia, and AMD.
\tool and \avclass do not show such drop.
\avclass is the fastest, followed by \tool, with \euphony being 
from 7 to 34 times slower than \avclass. 
Furthermore, we tested \euphony on the larger datasets without 
ground truth and for those larger than 1M samples, 
it did not terminate in 48 hours
or crashed with an out of memory exception 
(on a server with 126 GB RAM).

In summary, \avclass is more accurate and 7x--34x faster than 
\euphony. 
Furthermore, \euphony does not scale to large datasets due to its large 
memory usage.
{\avclass}* and \tool have the same accuracy, 
but \tool is slower due to the extra processing to extract more tags. 
Thus, there is no reason to use \tool over \avclass for family labeling, 
as long as both tools use the same input knowledge. 
Without providing updated data files to \avclass, 
\tool would outperform it, 
despite the family labeling functionality being identical.
This highlights the importance of the update module. 

An analyst that requires advanced searches or the extra intelligence should 
use \tool, but those that exclusively need family labels 
could still use the faster \avclass.
Our vision is that in the future both \tool and \avclass 
take as input the same data files, 
avoiding duplicate work to keep both tools up-to-date.
For this, we plan to integrate our {\avclass}* modifications into 
\avclass, so that both tools take as input the same files. 
This will enable the community to collaboratively update the malware knowledge 
in the taxonomy, tagging rules, and expansion rules, 
avoiding duplicate efforts, 
while maintaining backwards compatibility.

\section{Related Work}
\label{sec:related}

AV labels have been widely studied for over a decade.
Early works showed the problem of different AV engines disagreeing on 
labels for the same sample~\cite{bailey2007malware,canto2008large}.
Despite this problem, AV labels have been widely used to build 
training datasets and evaluate malware detection and clustering 
approaches~\cite{bailey2007malware,rieck2008learning,mcboost,bayer2009scalable,perdisci2010behavioral,malheur,bitshred,maggi2011finding,dahl2013large,drebin,malsign,miller16reviewer,mtnet}.
Li et al.~\cite{li2010challenges} studied AV labels as 
a reference to evaluate clustering results.
They showed that including only a subset of samples, 
for which AV engines largely agree, biases the evaluation towards 
samples that are easier to label.
Other works have focused on the quality of the labels from different 
AV engines. 
Mohaisen and Alrawi~\cite{avmeter} proposed metrics for evaluating AV labels, 
identifying clusters of AV engines that copy their labels~\cite{avmeter}; 
Kantchelian et al.~\cite{kantchelian2015better} 
discussed that AV engines have varying label quality and proposed
to weight them differently; and 
Hurier et al.~\cite{hurier16lack} proposed metrics to
evaluate ground truth datasets built using AV labels. 

The dynamics of AV labels have been another target of analysis.
Some works have shown how AV engines change their labels 
for the same samples over time as a result of signatures and 
analysis being refined~\cite{gashi13study,kantchelian2015better} 
and that detection systems should be trained with the labels available 
at training, not testing, time~\cite{miller16reviewer}. 
Recently, Zhu et al.~\cite{zhu20labels} analyze daily 
snapshots of 14K samples over a year. 
Among other results, they confirm that certain sets of AV engines produce 
strongly correlated labels, 
as observed in AVMeter~\cite{avmeter} and implemented in AVClass~\cite{avclass},
show that hand-picking of a few trusted engines does not always perform well, 
and measure that detection thresholds between two and 14 exhibit little 
differences in precision and recall.  

\paragraph{Malware labeling.}
One approach to tackle disagreements on malware names is 
to use naming conventions such as 
the 1991 CARO Virus Naming Convention~\cite{caro}.
Another attempt was the Common Malware Enumeration (CME) Initiative~\cite{cme} 
that provided unique identifiers for malware. 
Unfortunately, conventions have not achieved wide adoption, 
possibly due to their use of predefined tags, i.e., controlled vocabularies, 
that are incomplete and require frequent updates.

An alternative approach is to automatically 
extract accurate family names from AV labels. 
A precursor of this was VAMO~\cite{vamo} that proposed an automated 
approach for evaluating clustering results 
by building a graph of the normalized labels of 4 AV engines. 
It introduced the use of label co-occurrence,
i.e., the fraction of samples where labels, 
possibly from different engines, appear together. 

Tools like \avclass~\cite{avclass} and \euphony~\cite{euphony} 
have demonstrated that it is possible to extract accurate family tags
from AV labels.
One key idea of these works is to avoid using the whole label as family name,
as AV labels encode other non-family information.
\avclass and \euphony take as input AV labels
for a large number of samples, and output a family name for each sample.
They both use co-occurrence to automatically identify family aliases.
However, there are some key differences between them. 
\avclass proposes that aliases and generic tokens learned from one dataset 
can be reused on other datasets.
Instead, \euphony uses a graph-based approach that identifies aliases 
within the dataset, but does not produce relations that can be reused.
In addition, \avclass avoids predicting which AV engines are better 
at labeling samples.
Instead, it makes the tokenization as AV engine independent as possible. 
Instead, \euphony tries to learn the structure of the labels from a 
selected subset of AV engines.
As shown in Section~\ref{sec:eval_famlabels}, 
these differences result in improved accuracy and scalability for \avclass,
a key reason why we build \tool on top of \avclass.

Compared to prior malware labeling tools, \tool leverages that 
AV labels contain a wealth of information beyond family names 
such as malware classes, file properties, and behaviors. 
Such information is 
not extracted by tools like \avclass and \euphony
and can be used to produce tags to categorize and index malware samples. 

\paragraph{Malware tagging.}
The use of tags to characterize malware is at the heart of 
information sharing standards like 
Malware Attribute Enumeration and Characterization (MAEC)~\cite{maec} and 
Malware Information Sharing Platform (MISP)~\cite{mispStandard}. 
In addition, tags are already used by some malware repositories to enable 
efficient search. 
To further the use of malware tags,
we present \tool, an automatic tool to extract 
tags from the wealth of information on AV labels.
The generated tags can be incorporated by those repositories 
to enable richer searches.
Most related to our work is simultaneous work by 
Ducau et al.~\cite{ducau2019smart} that extract 11 pre-defined tags, 
mostly malware classes, from the AV labels of 10 AV engines. 
They use the tags as training set for a classifier that predicts the tags 
in new samples. 
Some key differences of \tool are that 
it does not pre-define the tags, instead building an open taxonomy 
(which currently has 150 non-family tags), 
handles tag aliasing 
(e.g., \TAG{downloader} and \TAG{dropper} are aliases in our taxonomy),
provides support for updating the input rules and taxonomy,
does not limit the supported AV engines, and 
is open source.

\section{Conclusion}
\label{sec:conclusion}

Automatically extracting tags from AV labels
is an efficient approach to categorize and index massive amounts of 
malware samples.
But, it is challenging due to the different vocabularies used by AV engines.
In this work, we have presented \tool, an 
open source automatic malware tagging tool.
Given the AV labels for a potentially massive number of samples,
\tool extracts clean tags that categorize the samples
according to their class, family, file properties, and behaviors.
The extracted tags can be used by malware repositories and analysis 
services to enable, or enhance, searches for samples of interest. 
Those samples, can in turn be used as ground truth for 
machine learning approaches.
\Tool uses, and helps building, an \emph{open taxonomy} that organizes
concepts in AV labels, but is not constrained to a predefined set of tags.
\Tool provides an update module that uses tag co-occurrence to automatically
identify taxonomy updates,
as well as tagging and expansion rules that capture relations between tags.
Thus, it can be easily updated as AV vendors introduce new tags.
We have evaluated \tool on \numsamples.

\begin{acks}
This research was supported by the Regional Government of Madrid 
through grants BLOQUES-CM P2018/TCS-4339 and PEJD-2018-PRE/TIC-9571
and by the Spanish Government through the SCUM grant RTI2018-102043-B-I00 and 
fellowship FPU18/06416.
Any opinions, findings, and conclusions or recommendations expressed in 
this material are those of the authors or originators, and 
do not necessarily reflect the views of the sponsors.

\end{acks}

\balance

\end{document}